\documentclass{iopart}
\usepackage{iopams,setstack,graphicx}
\begin{document}

\title{Influence of a Feshbach resonance on the photoassociation of LiCs}

\author{J Deiglmayr$^{1,2}$, P Pellegrini$^3$, A Grochola$^4$, M Repp$^1$, R C\^ot\'e$^3$, O Dulieu$^2$, R Wester$^4$ and M Weidem\"uller$^1$}
\address{$^1$ Physikalisches Institut, Ruprecht-Karls-Universit\"at Heidelberg, Germany}
\address{$^2$ Laboratoire Aim\'e Cotton, CNRS, Universit\'e Paris-Sud XI, Orsay, France}
\address{$^3$ Department of Physics, University of Connecticut, Storrs, USA}
\address{$^4$ Physikalisches Institut, Albert-Ludwigs-Universit\"at Freiburg, Germany}
\eads{\mailto{weidemueller@physi.uni-heidelberg.de},
\mailto{olivier.dulieu@lac.u-psud.fr} and
\mailto{rcote@phys.uconn.edu}}

\begin{abstract}
We analyse the formation of ultracold $^{7}$Li$^{133}$Cs molecules
in the rovibrational ground state through photoassociation into the
$B^1\Pi$ state, which has recently been reported [J. Deiglmayr
\textit{et al.}, Phys. Rev. Lett. \textbf{101}, 133004 (2008)].
Absolute rate constants for photoassociation at large detunings from
the atomic asymptote are determined and are found to be surprisingly
large. The photoassociation process is modeled using a full
coupled-channel calculation for the continuum state, taking all
relevant hyperfine states into account. The enhancement of the
photoassociation rate is found to be caused by an ``echo'' of the
triplet component in the singlet component of the scattering wave
function at the inner turning point of the lowest triplet
$a^3\Sigma^+$ potential. This perturbation can be ascribed to the
existence of a broad Feshbach resonance at low scattering energies.
Our results elucidate the important role of couplings in the
scattering wave function for the formation of deeply bound ground
state molecules via photoassociation.
\end{abstract}

\submitto{\NJP}
\maketitle

\section{Introduction}

%overview

The formation and manipulation of molecules at ultralow temperatures
has undergone tremendous progress in the past
years~\cite{doyle2004,dulieu2006}. The `holy grail' of the field is
the creation of a stable quantum-degenerate gas of molecules, in
which all molecules populate the ground state of the system,
comprising motional and internal degrees of freedom. Such a system
provides exquisite possibilities for further control and may thus
serve as the starting point to investigate future applications of
ultracold molecules such as precision measurements of fundamental
constants~\cite{zelevinsky2008} and ultracold chemical
reactions~\cite{tscherbul2006}. Dipolar molecules are of particular
interest in this context, as the dipolar interaction can be
exploited for the study of quantum many-body
phenomena~\cite{micheli2006, pupillo2008} or the realization of
various schemes for quantum information
processing~\cite{DeMille2002,yelin2006,rabl2006,charron2007}. The
most promising path towards the goal of a quantum-degenerate gas of
molecules in the internal ground state consists in the association
of ultracold atoms, either by magnetic or oscillatory electric
fields. In the first case, called magnetoassociation, the molecules
are formed through a magnetically induced Feshbach resonance
coupling the free pair of atoms to a weakly bound molecular state.
The transfer into the absolute internal ground state can then be
accomplished by stimulated rapid adiabatic passage (STIRAP) in a
combination of pulsed laser fields. This scheme has recently been
successfully applied to form deeply bound molecules of
Rb$_2$~\cite{lang2008}, Cs$_2$~\cite{danzl2008,danzl2008b}, and
KRb~\cite{ni2008}. The other, closely related approach consists of
the direct photoassociation of molecules out of an ultracold gas
followed by spontaneous emission, which may be followed by a second
bound-bound excitation step to transfer the molecules into the
vibrational ground state. This method has lead to the formation of
the vibrational ground state of K$_2$~\cite{nikolov2000},
RbCs~\cite{sage2005}, Cs$_2$~\cite{viteau2008}, and
LiCs~\cite{deiglmayr2008b}. Magnetoassociation combined with STIRAP
has the great advantage of being a fully coherent process which
preserves the phase-space density of the initial gas, while
photoassociation followed by spontaneous emission leads to smaller
phase-space densities but can be driven as a continuous process thus
allowing for the steady accumulation of molecules, e.g. in an
optical or static trap. Concerning ultracold dipolar gases we note
that among the above listed ultracold molecules in the ground state
the only dipolar ones are the heteronuclear molecules RbCs, KRb, and
LiCs. They all posses significant dipole moments of
1.2~\cite{aymar2005}, 0.57~\cite{ni2008}, and
5.5~Debye~\cite{aymar2005} respectively.

%%%%%%%%%%%%%%%%%%%%%%%%%%%%%%%%%%%%%%%%%%%%%%%%%%%%%%%%%%%%%%%%%%%%%%%%%%%%%%
\begin{figure}[tbh]
\begin{center}
\includegraphics[width=0.6\textwidth,clip]{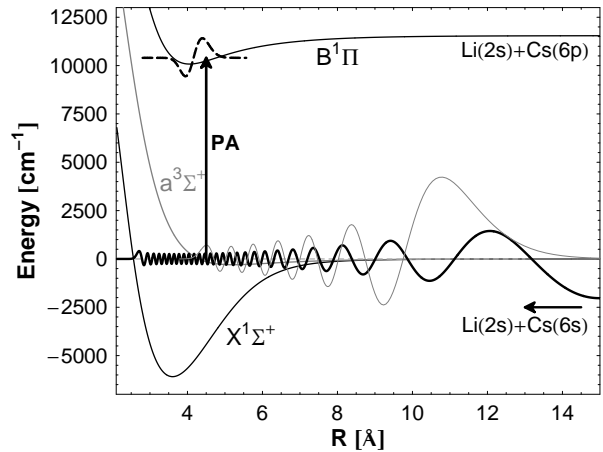}
\end{center}
\caption{\label{fig:intro} Schematic view of the photoassociation
process at short internuclear distances: the component of the
scattering wave function related to the singlet potential
$X^1\Sigma^+$ (solid black line) is coupled around the inner turning
point of the triplet potential $a^3\Sigma^+$ to a level in the
excited singlet state $B^1\Pi$ (black dashed line). A typical bound
state wave function in the $a^3\Sigma^+$ potential is also shown
(gray line).}
\end{figure}
%%%%%%%%%%%%%%%%%%%%%%%%%%%%%%%%%%%%%%%%%%%%%%%%%%%%%%%%%%%%%%%%%%%%%%%%%%%%%%

Here we focus on a recent experiment on the formation of ultracold
bosonic $^{7}$Li$^{133}$Cs ground state molecules with low
vibrational and rotational quantum
numbers~\cite{deiglmayr2008b,deiglmayr2009a}. The molecule formation
consists in a particularly simple two-step photoassociation (PA)
procedure comprising the laser excitation of the $B^1\Pi$ state out
of an ultracold mixture of lithium and cesium atoms. The vibrational
state $v'$=4 of this electronically excited state has a particularly
large Franck-Condon overlap with the vibrational level $v''$=0 of
the ground state $X^1\Sigma^+$ which leads to a measurable
population of this lowest vibrational level by spontaneous emission.
As only low angular momenta are involved due to the ultralow
temperature, ultracold molecules in the rovibrational ground state
could be detected. The rate limiting process of this PA scheme was
the excitation of the tightly bound $B^1\Pi$($v'$=4) level out of
the dilute gas. Actually, the observation of PA of molecules in this
state at short internuclear distance came as a surprise, as naive
estimates delivered negligible formation rates. In this article, we
investigate the PA process in more detail. We find that the PA rate
is strongly enhanced by an increased amplitude of the scattering
wave function at short internuclear distances. It is important to
note that despite the presence of spin-orbit coupling close to the
atomic asymptote, for deeply bound levels the excited state has very
pure singlet character. Therefore only the singlet component of the
scattering wave function is relevant for the PA rate.
Figure~\ref{fig:intro} depicts this situation: the Franck-Condon
overlap between the scattering wave function in the singlet
potential $X^1\Sigma^+$ and low lying levels in the $B^1\Pi$
potential vanishes, as the fast oscillations of the scattering wave
function at short internuclear distances average to zero. However,
experimentally we observe high PA rates into excited levels, which
correspond to a transition around the inner turning point of the
$a^3\Sigma^+$ state. Therefore we conclude that the scattering wave
function is locally perturbed by the presence of a bound level with
strong triplet character, indicating the proximity of a Feshbach
resonance. Indeed the importance of closed coupled channels for the
scattering wave function close to a Feshbach resonance has recently
been theoretically explored by some of us as a means to strongly
enhance the PA formation rate of ultracold
molecules~\cite{pellegrini2008}. The perturbation introduced by this
Feshbach resonance leads to an ``echo'' of the triplet-like wave
function on the singlet component of the scattering wave function,
which increases the Franck-Condon overlap between the scattering
wave function and tightly bound levels in the $B^1\Pi$ potential. We
present a model containing an accurate and complete description of
the continuum scattering wave function and of the excited molecular
wave function, which proves the validity of this picture and
reproduces the experimental observations accurately.

The paper is organized as follows: In section~\ref{sec:formation} we
describe our experiment on the formation and detection of ultracold
LiCs molecules in low vibrational and rotational states of the
singlet ground state. Section~\ref{sec:linestrengths} is then
devoted to a quantitative analysis of measured line strengths and
rate constants for PA into the $B^1\Pi$ state. In
section~\ref{sec:model} we develop the theoretical model: A brief
overview is given on how PA rates are computed based on wave
functions of the excited levels. It is explained, how hyperfine
interactions in the scattering entrance channel couple singlet and
triplet components and lead to an ``echo'' of triplet features in
the singlet component of the scattering wave function. Numerical
methods are introduced, and the results obtained by the
coupled-channel calculation are discussed. Results for the coupled
and uncoupled cases are compared with the experimental findings.
Section~\ref{sec:conclusion} discusses the results of the paper.

\section{Formation and detection of ultracold LiCs molecules}
\label{sec:formation}

% Trap setup
Details of the experimental setup for the formation and detection of
ultracold LiCs molecules have already been described in detail in
Refs.~\cite{deiglmayr2008b,deiglmayr2009a,kraft2007}. Therefore, we
will confine ourselves to the description of the main experimental
features.  $4\times10^7$ $^{133}\,$Cs atoms and 10$^8$ $^{7}$Li
atoms are trapped in overlapped magneto-optical traps (MOT's) at
densities of $3\times10^9\,$cm$^{-3}$ and $10^{10}\,$cm$^{-3}$
respectively. The MOT for cesium is realized as a dark spontaneous
force optical trap \cite{ketterle1993}, in which the atoms are kept
most of the time (typically 97\%) in the dark lower hyperfine ground
state leading to higher densities and to a reduction of inelastic
collisions. In the conventional lithium MOT, most of the atoms
populate the upper hyperfine ground state ($>$80\%). Lithium and
cesium atoms therefore collide mainly on the
Li(2$^2$S$_{1/2}$,$f$=2)+Cs(6$^2$S$_{1/2}$,$f$=3) asymptote. We
measure a cesium temperature of 250(50)$\mu$K using time-of-flight
expansion and deduce a lithium temperature of 600(150)$\mu$K by
fitting the line shape of a narrow PA
resonance~\cite{deiglmayr2009a} to the model of
reference~\cite{jones1999}. In the center-of-mass frame, this
corresponds to a mean collision energy of 580(80)$\mu$K.

\begin{figure}[htb]
\begin{center}
\includegraphics[width=0.48\textwidth,clip]{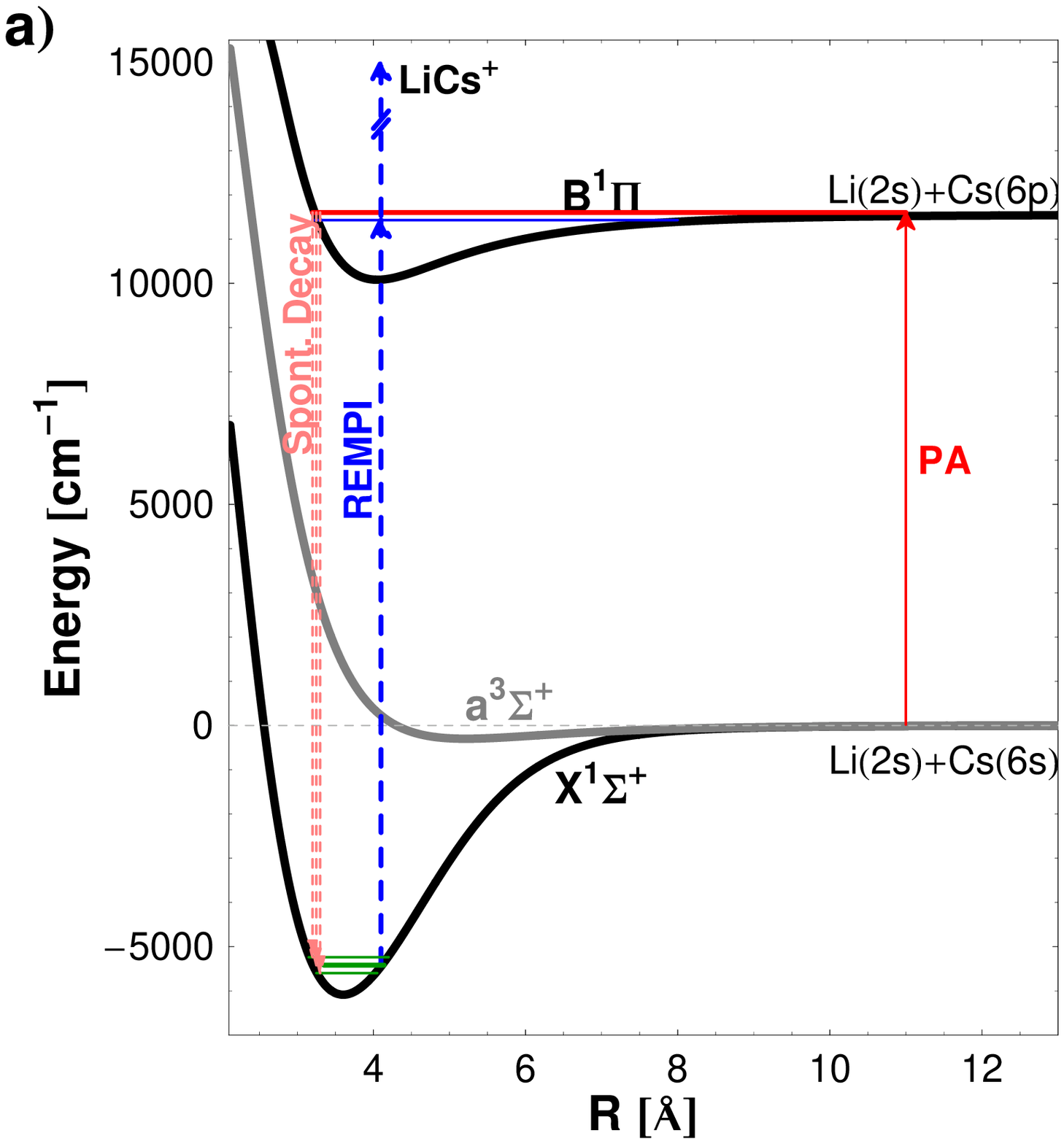}
\includegraphics[width=0.48\textwidth,clip]{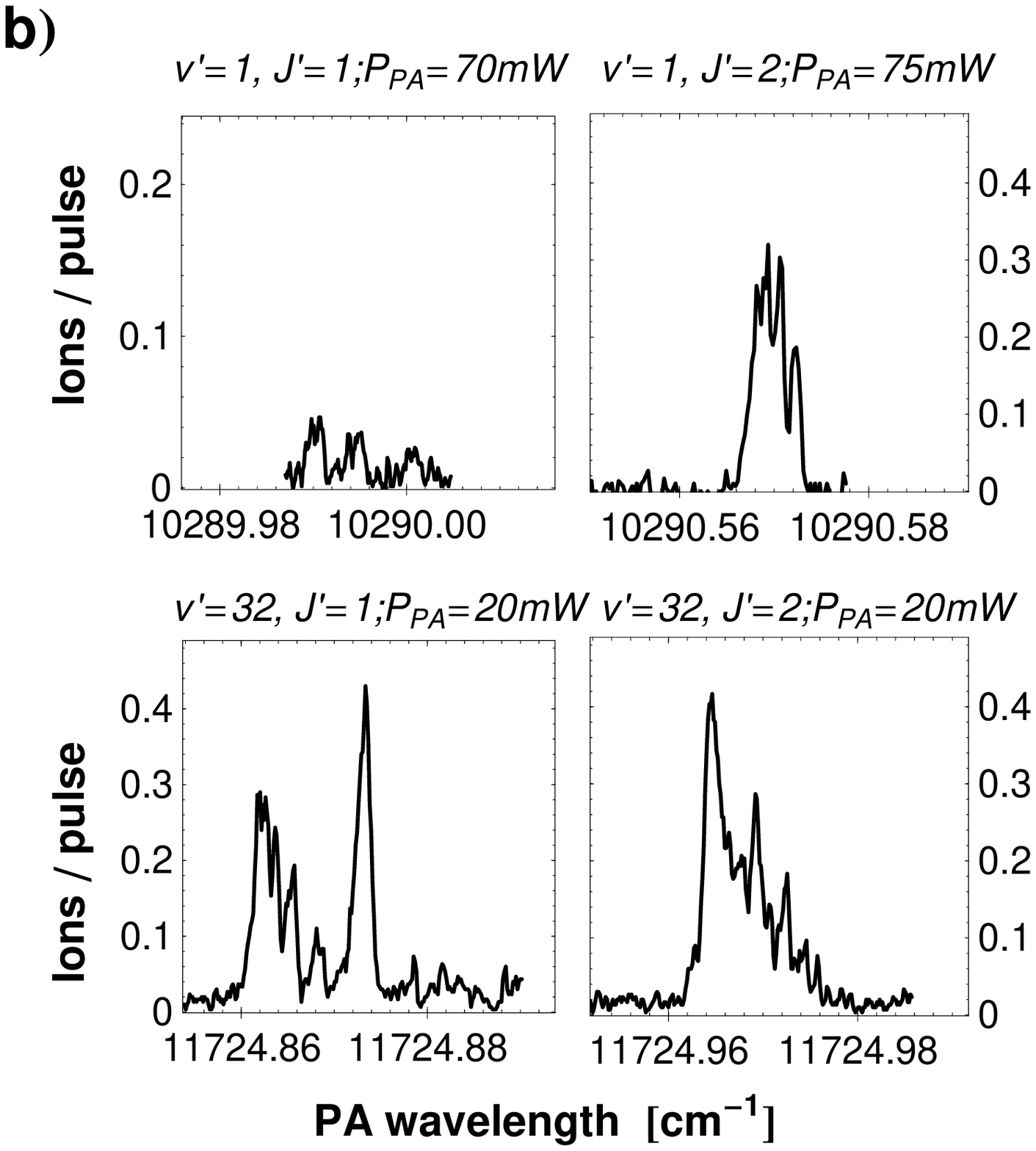}
\end{center}
\caption{\label{fig:pascheme} (a) Photoassociation and detection
scheme together with the relevant levels of LiCs: after PA into
levels of the $B^1\Pi$ state, molecules decay spontaneously into
$X^1\Sigma^+$ molecules. These molecules are ionized by two photons,
where resonant intermediate states enhance the ionization rate. (b)
Exemplary photoassociation resonances for deeply bound levels
($v'$=1) and levels close to the asymptote ($v'$=31) (detected at
different REMPI wavelengths, see text).}
\end{figure}

% PA and detection
In PA, a colliding pair of atoms absorbs a photon resonant to a
transition into a bound excited molecular
level~\cite{thorsheim1987,jones2006}. For this process we use up to
500\,mW of light from a tunable Ti:Sa laser. The beam is collimated
to a waist of 1.0\,mm and continuously illuminates the two
overlapped atom traps. The excited molecules decay within a few tens
of nanoseconds either into bound ground state molecules or back into
free pairs of atoms with additional kinetic energy~\cite{cote1997}.
For the detection of ground state molecules, we use a pulsed dye
laser with a repetition rate of 20\,Hz (typical pulse energy 8\,mJ,
beam diameter $\sim$5\,mm). Molecules in the ground state are
ionized by resonant-enhanced multi-photon ionization (REMPI), in our
case by two photons of the same color. The resulting LiCs$^+$ ions
are then separated from a background of Cs$^+$ ions in a
time-of-flight mass spectrometer and are finally detected using a
micro-channel plate and single-ion-counting electronics.
Figure~\ref{fig:pascheme}\,(a) depicts the states involved in the
formation and detection of ground state molecules. Exemplary PA
resonances are shown in figure~\ref{fig:pascheme}\,(b).

%%%%%%%%%%%%%%%%%%%%%%%%%%%%%%%%%%%%%%%%%%%%%%%%%%%%%%%%%%%%%%%%%%%%%%%%%%%%%%
%%%%%%%%%%%%%%%%%%%%%%%%%%%%%%%%%%%%%%%%%%%%%%%%%%%%%%%%%%%%%%%%%%%%%%%%%%%%%%

% The B1Pi
We focus on the $B^1\Pi$ state of LiCs, which has been studied
previously in a heat-pipe with high-resolution laser-induced
fluorescence spectroscopy~\cite{stein2008}. Due to spin-orbit
interaction, this state is asymptotically correlated to
Li(2$^2$S$_{1/2}$)+Cs(6$^2$P$_{3/2}$) after its coupling with
neighboring triplet states. For convenience, we will refer to it as
the $B^1\Pi$ state. Additionally to the vibrational levels $v'=0-25$
identified in reference~\cite{stein2008}, we found vibrational
levels $v'=26-35$, where the last level $v'=35$ is bound only by a
few GHz. For most lines, we observe rotational components $J'=1$ and
$J'=2$, which are split into several hyperfine sub-lines. The
typical overall width of these hyperfine structures ranges for
$J'=1$ resonances from 550\,MHz for low vibrational levels to
900\,MHz for the last bound levels, while the $J'=2$ resonances are
narrower, ranging from a width of 300\,MHz for low $v'$s to 600\,MHz
close to the asymptote.

%Detection
We detect PA resonances by ionozing ground state molecules produced
after spontaneous decay from the photoassociated, electronically
excited molecules. Three different frequencies were used for the
ionization of these ground state molecules. Molecules produced by PA
in low vibrational states $v'$$<$9 are ionized at
16999.7\,cm$^{-1}$, where three resonant transitions for the first
step, $X^1\Sigma^+$,$v''$=0$\rightarrow$$B^1\Pi$,$v'$=14,
$X^1\Sigma^+$,$v''$=1$\rightarrow$$B^1\Pi$,$v'$=18, and
$X^1\Sigma^+$,$v''$=2$\rightarrow$$B^1\Pi$,$v'$=23, are nearly
degenerate~\cite{deiglmayr2009a}. Ground state molecules formed
after PA via intermediate vibrational levels $v'$$\geq9$ and
$v'$$\leq15$ are detected at 16859.4\,cm$^{-1}$, a REMPI resonance
where a strong contribution from the transition
$X^1\Sigma^+$,$v''$=2$\rightarrow$$B^1\Pi$,$v'=19$ was
identified~\cite{deiglmayr2009a}. While the here identified ground
state levels make up for an important part of the observed ion
signal, molecules in other ground state levels might also be ionized
by the given laser frequency via intermediate states other than
$B^1\Pi$. Molecules which have decayed from high lying vibrational
levels $v'>15$ are all ionized at 14692.7\,cm$^{-1}$. The ionization
process at this resonance has not yet been analysed. However, it is
very likely that high lying levels in the electronic ground state
$X^1\Sigma^+$ are ionized at this energy through a broad band of
levels in different electronic states between the asymptotes
Li(2$^2$S$_{1/2}$)+Cs(6$^2$P$_{3/2}$) and
Li(2$^2$S$_{1/2}$)+Cs(5$^2$D$_{3/2}$).

\section{Line strengths of the photoassociation resonances and formation rate constants}
\label{sec:linestrengths}

\subsection{Photoassociation resonance line strengths}

%%%%%%%%%%%%%%%%%%%%%%%%%%%%%%%%%%%%%%%%%%%%%%%%%%%%%%%%%%%%%%%%%%%%%%%%%%%%%%
\begin{figure}[htb]
\begin{center}
\includegraphics[width=0.7\textwidth,clip]{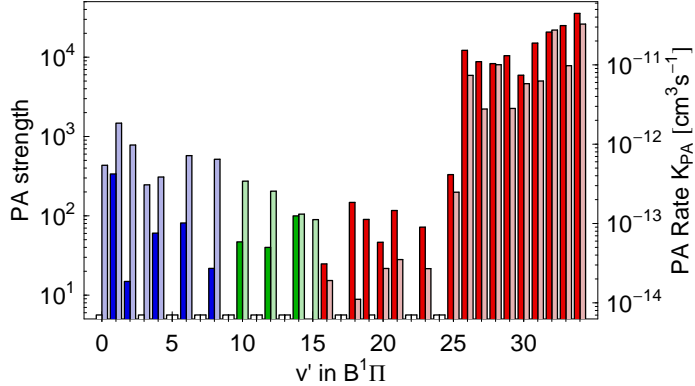}
\end{center}
\caption{\label{fig:pastrength} Photoassociation line strength
(given as number of molecules$\times$MHz/s/mW) and PA rates of all
observed levels $v'$,$J'$=1 (left, dark colors) and $J'$=2 (right,
light colors) in the $B^1\Pi$ state. The PA resonances are
integrated and normalized to the PA laser power. Absolute rates are
given for a reference PA intensity of 30W/cm$^2$. Ionization
wavelengths used for detection are 14692.7\,cm$^{-1}$ (red),
16859.4\,cm$^{-1}$ (green), and 16999.7\,cm$^{-1}$ (blue),
respectively.}
\end{figure}
%%%%%%%%%%%%%%%%%%%%%%%%%%%%%%%%%%%%%%%%%%%%%%%%%%%%%%%%%%%%%%%%%%%%%%%%%%%%%%

%The PA line strengths
The strength of the PA resonances are shown in
figure~\ref{fig:pastrength} for all observed $v'$, $J'$=1 and $J'$=2
levels. The last bound level $v'$=35 was excluded from the analysis,
as the cesium MOT is strongly perturbed by the PA laser at small
detunings from the atomic asymptote. In order to reduce the
influence of the varying hyperfine structure, the full PA spectrum
is background-subtracted and integrated for each ro-vibrational
level, therefore summing up the contributions from all hyperfine
components. In order to directly compare the line strength of PA
resonances measured under different experimental conditions, the
integrated peak area is normalized to the intensity of both the PA
laser and the ionization laser. At a strong line we confirmed for
both the PA and the REMPI step that the molecule formation rate is
indeed linear in both intensities. For the two-photon ionization,
this indicates that the first resonant-enhanced transition is
strongly saturated and therefore the probability for this transition
does not depend on the intensity of the ionization laser above a
certain, not determined threshold. The PA laser intensities were
chosen for each resonance in such a way that the maximum count rate
does not exceed 0.5\,ions/pulse to avoid saturation of the single
ion detection.

One uncertainty remains, which limits the comparability between
different PA resonances. As discussed in Sect.~\ref{sec:formation},
we do not detect the excited molecules directly but only the ground
state molecules formed after spontaneous decay. As the Frank-Condon
factors between excited state and ground state levels vary with the
PA level $v'$, also the distribution of populated ground state
levels after spontaneous decay depends on the PA
level~\cite{deiglmayr2009a}. These differences in the populated
levels can lead to variations in the overall ionization efficiency
for molecules produced at different PA resonances. It is difficult
to estimate the influence of this varying detection efficiency on
the measured PA strengths, as many off-resonant and nonlinear
processes play a role in the REMPI ionization scheme. We observe
however that the calculated ground state distributions vary only
slowly with the excited level $v'$. Also from measurements of the
same PA line at different ionization resonances, we deduce that this
variation does not exceed a factor of ten over neighboring
vibrational levels. Nonetheless this constitutes one of the main
uncertainties in the reported PA rate constants. For levels close to
the asymptote we also compare the molecule formation rate deduced
from the ion signal with measurements of trap loss. We find fair
agreement within one order of magnitude, which is acceptable in view
of the large systematic uncertainty on the efficiency of the ion
detection.

\subsection{Absolute molecule formation rates}
\label{sec:exp_form_rate}

% Detection efficiency
In order to relate the  PA line strengths to the formation rate of
molecules, we further analyse the detection efficiency for ground
state molecules. As the formed molecules are not trapped, they leave
the ionization region some time after their production due to their
thermal velocity of roughly 300\,$\mu$K (calculated from the
velocities of the atomic constituents) and acceleration due to
gravity. Taking the size and the alignment of the ionization beam
(roughly one beam diameter below the trapped atom clouds) into
account, we estimate a geometric overlap factor of 40\%. Therefore,
only 40\% of the formed molecules have a chance to be ionized.
Assuming that the first bound-bound transition during the ionization
of ground state molecules is on resonance and thus fully saturated,
we use a typical ionization cross section of
10$^{-18}$cm$^{2}$~\cite{verner1996} to approximate the ionization
probability $p_{\mathrm{ion}}$ for pulse energies
E$_{\mathrm{REMPI}}$ of up to few tens of mJ by
$p_{\mathrm{ion}}$\,$\simeq$\,8$\times$10$^{-3}$E$_{\mathrm{REMPI}}$
where E$_{\mathrm{REMPI}}$ is measured in mJ. Finally the detector
efficiency is around 20\%~\cite{fraser2002}, leading to an overall
efficiency of the detection setup on the order of 10$^{-3}$ for
typical experimental parameters. By averaging over many cycles
(typically around 100), even very weak PA resonances can be
detected.

The rate constant $K_{\mathrm{PA}}$ for the PA process is calculated
by first linearly scaling the number of produced molecules per
second to a reference PA intensity (chosen to be 30\,W/cm$^2$)
yielding a molecular formation rate $k$. In our experimental
geometry the diameter of the PA laser beam and of the lithium MOT
are both larger than the size of the cesium MOT. The rate
coefficient $K_{\mathrm{PA}}$ is then simply given by
$K_{\mathrm{PA}}=k/n_{Li}/N_{Cs}$ with the peak lithium density
$n_{Li}$ and the cesium particle number $N_{Cs}$. The measured rate
coefficients are shown in figure~\ref{fig:pastrength} (right axis).
For the last bound states we observe PA rate constants on the order
of 4$\times$10$^{-11}$cm$^3$/s. These are of the same order of
magnitude as the rate constants observed in the PA of
Cs$_2$~\cite{drag2000,wester2004} and KRb~\cite{mancini2004a}, as
was predicted by Azizi \textit{et al.}~\cite{Azizi2004}.

The PA rate decreases towards lower vibrational states, which are
excited at shorter internuclear distance corresponding to lower pair
density. Below $v'=$25 we observe a change towards a distinct
oscillatory pattern in the rate constants. While some vibrational
levels show rate constants varying by no more than one order of
magnitude, adjacent levels are below the detection threshold of our
setup. For very deeply bound levels $v'$$\leq$6 the PA rate appears
to increase again, \textit{e.g.} for the $v'$=1,$J'$=1 level at a
detuning of more than 1.400\,cm$^{-1}$ we measure a PA rate
coefficient on the order of 4$\times$10$^{-13}$cm$^3$/s. This is
unexpectedly high, as a single channel scattering calculation
predicts a rate constant below 10$^{-19}$cm$^3$/s due to the low
Franck-Condon overlap between the rapidly oscillating singlet
scattering wave function and the excited state wave function at
short internuclear distances. In the following, we describe the
theoretical framework employed to explain the unexpectedly large
formation rate in the singlet electronic state.

\section{Model of photoassociation into the $B^1\Pi$ electronic state}
\label{sec:model}

\subsection{Theoretical photoassociation rates}

A PA rate $K_{\rm PA}^{v'}=\langle v_{\rm rel}
\sigma_{PA}^{v'}\rangle$ from an initial scattering continuum state
to an excited vibrational level $v'$ is obtained using the cross
section $\sigma_{PA}^{v'}$ of two atoms colliding in the presence of
a laser field, and averaged over the relative velocity distribution
$v_{\rm rel}$. For a thermal atomic gas, a Maxwell-Boltzmann
distribution characterized by a temperature $T$ is appropriate, and
assuming a PA laser beam of negligible linewidth and resonant for
the transition to $v'$, we have \cite{napolitano1994}:
\begin{equation}
\label{rate_eq}
    K_{\rm PA}^{v'}(T,I)=\frac{1}{hQ_T}\int_0^{\infty}d\varepsilon \;e^{-\varepsilon/k_BT}
   \frac{\gamma_{v'} \gamma_s}{\varepsilon^2+(\frac{\gamma_{v'}+\gamma_s}{2})^2} ,
\end{equation}
where $Q_T=(2 \pi \mu k_BT/h^2)^{3/2}$, $h$ and $k_B$ are the Planck
and Boltzmann constant, respectively, $\varepsilon = \mu v^2_{\rm
rel}/2$ is the relative kinetic energy of the colliding pair of
atoms of reduced mass $\mu$, and $\gamma_{v'}$ is the natural
linewidth of the photoassociated level.

Using Fermi's Golden Rule, we obtain an expression for the
stimulated emission width $\gamma_s$ in terms of the initial
scattering state $\Psi_{\epsilon,l}$ at energy $\epsilon$, and the
excited vibrational state $\phi_{v',J'}$
\begin{equation}
 \gamma_s=\frac{\pi I}{\epsilon_0 c}|\langle\phi_{v',J'}|D(R)|\Psi_{\epsilon,\ell}\rangle|^2 .
 \label{eq:gamma_s}
\end{equation}
Here, $\ell$ labels the partial wave, $I$ is the intensity of the PA
laser, and $\epsilon_0$ and $c$ are the vacuum permittivity and
speed of light, respectively. We will restrict ourselves to $s$-wave
scattering in the entrance channel, and thus set $\ell$=0. This is
well justified, as the measured relative temperature in the
experiment of 580\,$\mu$K is well below the $p$-wave barrier height
of 1.6\,mK.

We obtain the scattering wave function in the entrance channel
$\Psi_{\epsilon,l}$ by solving the Hamiltonian of two colliding
atoms with hyperfine interactions, which mix the singlet
$X^1\Sigma^+$ and triplet $a^3\Sigma^+$ states. In addition to this
scattering wave function and the wavefunction of the excited level
$\phi_{v',J'}$ (see descriptions in the next sections), we need the
dipole transition moment $D(R)$ to calculate $\gamma_s$. Its
$R$-dependent form is calculated using the theoretical method and
parameters described in reference~\cite{aymar2005}. The value of the
dipole moment does not vary by more than 10\% around a mean value of
4.0\,a.u. over the relevant range of distances.

\subsection{Excited state vibrational wave functions}
\label{sec:excited_wfcts}

PA is performed to the excited $B^1\Pi$ molecular potential. High
resolution spectroscopy of this potential has been published in
reference~\cite{stein2008}. We use a potential fitted to the levels
observed in this work combined with the additional levels observed
by us~\cite{inprepA}. However, we note that the spin-orbit coupling
of the $B^1\Pi$, $b^3\Pi$, and $c^3\Sigma^+$ states becomes
important near the dissociation limit and cannot be neglected. The
Hund's case (a) is therefore no longer valid and the $B^1\Pi$ state
should be treated as $\Omega$=1 in Hund's case (c). To estimate the
strength of the singlet-triplet mixing we used a simple 3-channel
model~\cite{bergeman2002}, coupling ab-initio curves for $B^1\Pi$,
$b^3\Pi$, and $c^3\Sigma^+$ (calculated as described in
reference~\cite{aymar2005}) with constants proportional to the
atomic finestructure interaction of
554.04\,cm$^{-1}$~\cite{udem1999,udem2000} for the first asymptote
of cesium. Wave functions and projections on the three channels
where calculated using the mapped Fourier grid
method~\cite{kokoouline1999}.

%%%%%%%%%%%%%%%%%%%%%%%%%%%%%%%%%%%%%%%%%%%%%%%%%%%%%%%%%%%%%%%%%%%%%%%%%%%%%%
\begin{figure}[tb]
\begin{center}
\includegraphics[width=0.7\textwidth,clip]{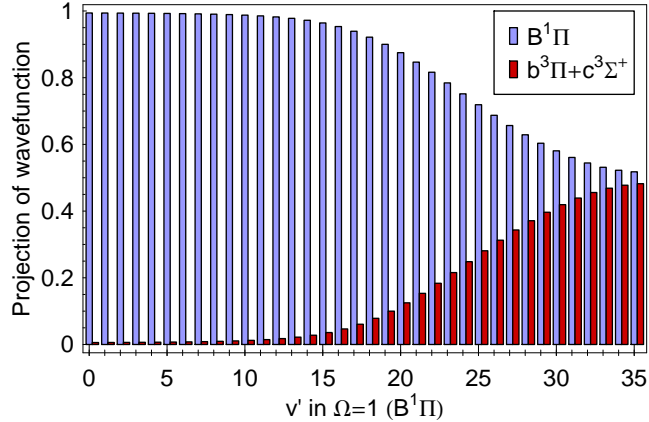}
\end{center}
\caption{Projection of the $\Omega$=1\,($B^1\Pi$) wave functions
onto the singlet component $B^1\Pi$ (light blue) and the triplet
components $b^3\Pi$ and $c^3\Sigma^+$ (dark red). At low $v'$s the
state can therefore be identified as pure $B^1\Pi$, while for
$v'$$\sim$20 and higher the triplet character, induced by spin-orbit
coupling, becomes important.} \label{fig:SO}
\end{figure}
%%%%%%%%%%%%%%%%%%%%%%%%%%%%%%%%%%%%%%%%%%%%%%%%%%%%%%%%%%%%%%%%%%%%%%%%%%%%%%

As can be seen in figure~\ref{fig:SO}, the components of the
$\Omega$=1\,($B^1\Pi$) wave function related to the coupled triplet
states become sizable above $v'=20$. For the highest lying levels,
the summed amplitude of these components reaches nearly 50\%. This
value can be taken as a first estimate for the influence of the
excited state mixing on the PA line strength: the triplet component
of the excited state can now be populated from the triplet component
of the scattering state. In general (\textit{i.e.} as long as there
are no strong local perturbations of the wave functions), the
excited triplet components will only decay into triplet ground state
molecules or free pairs of atoms, so they will not contribute to the
production of deeply bound singlet molecules. In such a case one
would expect for high lying levels to observe PA rates, which are
reduced by the amount of the triplet projection shown in
figure~\ref{fig:SO} when compared to PA rates derived from the here
presented theory. However, since the focus of this work is on the
lowest vibrational levels (below $v'=20$), we treated the excited
potential as a pure $B^1\Pi$ state in our calculation.

\subsection{Scattering wave function of the ground state}

If we label $\vec{s}_j$ and $\vec{i}_j$ the electronic and nuclear
spin for the atom $j$, respectively, then the total spin $f_j$ of
atom $j$ (with projection $m_{f_j}$) is given by
$\vec{f}_j=\vec{s}_j+\vec{i}_j$. The total angular momentum without
rotation is $\vec{F}=\vec{f}_1+\vec{f}_2$. The two-body hyperfine
Hamiltonian for atoms of relative momentum $\vec{p}$ (and reduced
mass $\mu$) can be written as
\begin{equation}
  \label{Hamiltonian}
  H=\frac{p^2}{2\mu}+\sum_{j=1}^{2} \frac{a^{(j)}_{\rm hf}}{\hbar^2}\vec{s}_j\cdot\vec{i}_j
  +V^c \; ,
\end{equation}
where $V^c$ is the potential energy operator, and $a^{(j)}_{\rm hf}$
is the hyperfine constant for atom $j$. No spin-rotation coupling is
introduced here, which means that the rotational angular momentum
$\ell$ is fixed. $V^c$ can be expressed in terms of the singlet
$V_0(R)$ and triplet $V_1(R)$ molecular potentials using the
projection operators $P^0$ and $P^1$:
\begin{equation}
  V^c=V_0(R)P^0+V_1(R)P^1 \;.
\end{equation}
As it is the case for the excited $B^1\Pi$ state, the ground
electronic singlet and triplet potentials are known from high
resolution spectroscopy \cite{staanum2007}. We used the potential
curves as given in the reference, with a cubic spline interpolation
of the given pointwise representation connected to the given
long-range expansion.

The hyperfine Hamiltonian in Eq.(\ref{Hamiltonian}) only couples
hyperfine channels with the same total angular momentum projection
$m_F=m_{f_1}+m_{f_2}$. For $^7$Li with $i_1=3/2$ and $^{133}$Cs with
$i_2=7/2$ (both with $s_1=s_2=1/2$), $m_F$ takes values from -6 to
+6, leading to a total degeneracy of the atom pair of 128. However,
atoms mainly collide on the
Li$(2^2S_{1/2},f=2)$+Cs$(6^2S_{1/2},f=3)$ asymptote, and thus only
35 degenerate entrance channels have to be considered.

The hyperfine Hamiltonian (\ref{Hamiltonian}) is expressed and
diagonalized in the atomic basis
\begin{equation}
  \label{Vdd2}
  |f_1,m_{1};f_2,m_{2}\rangle\equiv|f_1,m_{1}\rangle_{\rm Li}\otimes
  |f_2,m_{2}\rangle_{\rm Cs} \;,
\end{equation}
and the total collisional wave function is then expressed as
\begin{equation}
\label{wf1}
 |\Psi_{\epsilon,\ell}\rangle=\sum_{\alpha=1}^{N} \psi_{\alpha}(R)
 \lbrace | f_1,m_{1}\rangle\otimes|f_2,m_{2}\rangle\rbrace_{\alpha} \; ,
\end{equation}
where $\alpha$ labels a particular channel, and $N$ is the number of
coupled hyperfine channels which depends on the value of $m_F$.
Here, $\psi_{\alpha}(R)$ is the radial wave function of channel
$\alpha$.

For the calculation of the Franck-Condon factors, only the singlet
component of the total collisional wave function is relevant. A
rotation is thus performed to express the wave function in the
molecular basis $|S, m_S,I,m_I\rangle$ with
$\vec{S}=\vec{s}_1+\vec{s}_2$ and $\vec{I}=\vec{i}_1+\vec{i}_2$. We
then have
\begin{equation}
 |\Psi_{\epsilon,\ell}\rangle=\sum_{\beta=1}^{N} \phi_{\beta}(R)
 \lbrace | S,m_S,I,m_I\rangle\rbrace_{\beta} \; ,
\end{equation}
where $\phi_{\beta}(R)$ is the wave function in this basis (with $N$
channels $\beta$). We can go from one basis to another basis using
angular momentum algebra.

%\begin{eqnarray*}
%|f_1,m_{f_1};f_2,m_{f_2}\rangle=
%\sum_{F,m_F,I.m_I,S,m_S}(-1)^{\underline{}f2-f1-m_F+S-I-m_F}(2F+1)\\
%\times \sqrt{(2f_1+1)(2f_2+1)(2I+1)(2S+1)}\\
%\times\left(\begin{array}{ccc}f_1&f_2&F \\
%m_{f_1}&m_{f_2}&-m_{F}\end{array}\right)
%\left(\begin{array}{ccc}I&S&F \\
%m_{I}&m_{S}&-m_{F}\end{array}\right)
%\left\{ \begin{array}{ccc} i_1&s_1&f_1\\
%i_2&s_2&f_2 \\ I&S&F \end{array} \right\}| S m_S, I m_I\rangle
%\textit{}\end{eqnarray*}

\subsection{Results for $m_F=5$ scattering state}

%%%%%%%%%%%%%%%%%%%%%%%%%%%%%%%%%%%%%%%%%%%%%%%%%%%%%%%%%%%%%%%%%%%%%%%%%%%%%%
\begin{table}[b]
\begin{center}
\begin{tabular}{c|c|c}
&Atomic basis&Molecular basis\\
\hline
$m_F=5$&$f_1=2,m_{f1}=2,f_2=3,m_{f2}=3$&$S=0,m_S=0,I=5,m_I=5$\\
       &$f_1=2,m_{f1}=2,f_2=4,m_{f2}=3$&$S=1,m_S=0,I=5,m_I=5$\\
       &$f_1=1,m_{f1}=1,f_2=4,m_{f2}=4$&$S=1,m_S=1,I=4,m_I=4$\\
       &$f_1=2,m_{f1}=1,f_2=4,m_{f2}=4$&$S=1,m_S=1,I=5,m_I=4$\\
\end{tabular}
\end{center}
\caption{\label{tab:mf=5} Values for the quantum numbers of the
collisional channels of the $m_F=5$ subspace for the atomic basis
$\left|f_1,m_{f1},f_2,m_{f2}\right>$ and the molecular basis
$\left|S, m_S,I,m_I\right\rangle$. Note that there is no
line-to-line relation between the states listed in each column.}
\end{table}
%%%%%%%%%%%%%%%%%%%%%%%%%%%%%%%%%%%%%%%%%%%%%%%%%%%%%%%%%%%%%%%%%%%%%%%%%%%%%%

%%%%%%%%%%%%%%%%%%%%%%%%%%%%%%%%%%%%%%%%%%%%%%%%%%%%%%%%%%%%%%%%%%%%%%%%%%%%%%
\begin{figure}[t]
\begin{center}
\includegraphics[width=0.8\textwidth,clip]{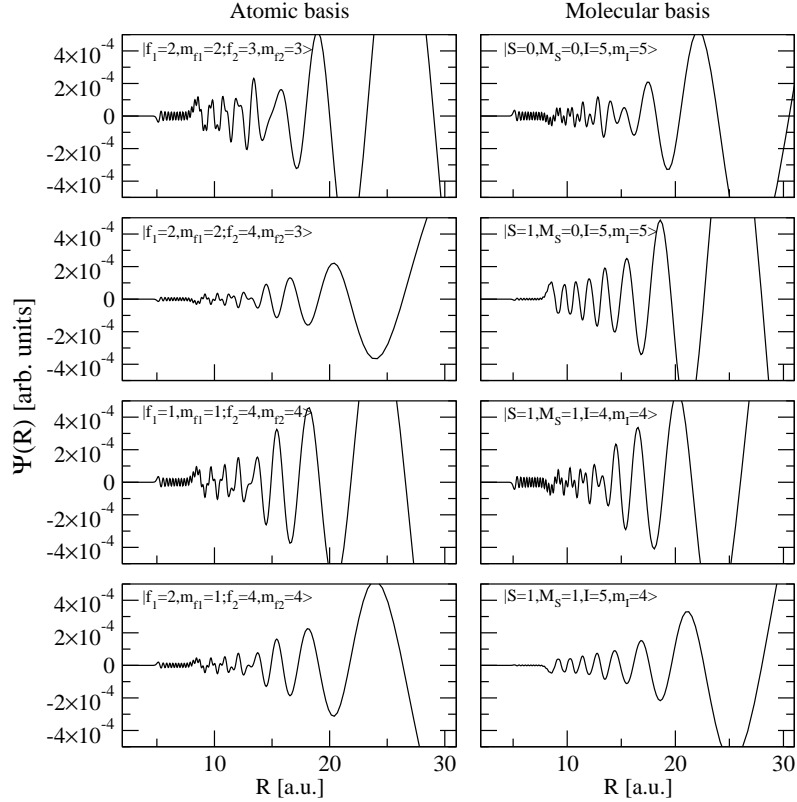}
\end{center}
\caption{\label{fig:wf_mf=5} Components of the energy-normalized
radial wave function for a Li-Cs pair colliding at 500~$\mu$K, in
the atomic basis (left columns) and in the molecular basis (right
column). Under the given experimental conditions, the upper-most
atomic state $f_1$=2,$m_{f1}$=2,$f_2$=3,$m_{f2}$=3 is the only open
entrance channel, the three other shown states are closed channels.
Note that there is no line-to-line relation between the states
listed in each column.}
\end{figure}
%%%%%%%%%%%%%%%%%%%%%%%%%%%%%%%%%%%%%%%%%%%%%%%%%%%%%%%%%%%%%%%%%%%%%%%%%%%%%%

%%%%%%%%%%%%%%%%%%%%%%%%%%%%%%%%%%%%%%%%%%%%%%%%%%%%%%%%%%%%%%%%%%%%%%%%%%%%%%
\begin{figure}[tb]
\begin{center}
\includegraphics[width=0.6\textwidth,clip]{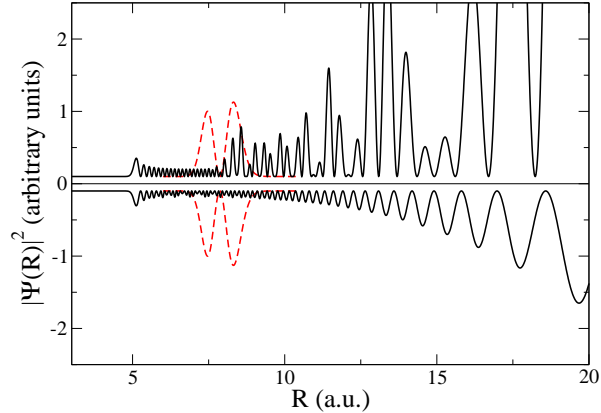}
\end{center}
\caption{\label{fig:wf_mf=5-B} Black lines: Singlet projection of
the total wave function in the coupled-channel model (upper panel),
and in a single channel calculation with the singlet potential
(lower pannel). Probability densities for the $v'$=1 level of the
$B^1\Pi$ state are also displayed (red dashed lines), illustrating
the increased overlap due to the enhancement of the ground state
wave function at the inner turning point of the triplet potential.}
\end{figure}
%%%%%%%%%%%%%%%%%%%%%%%%%%%%%%%%%%%%%%%%%%%%%%%%%%%%%%%%%%%%%%%%%%%%%%%%%%%%%%

For simplicity we concentrate here on the particular case of
$m_F=5$, corresponding to the coupling of the four channels listed
in Table \ref{tab:mf=5} by the hyperfine interaction. These four
states represent approximately 1/10th of a total of 35 degenerate
channels of the entrance channel. In order to compute the dipole
transition matrix elements
$\langle\phi_{v',J'}|D(R)|\Psi_{\epsilon,l}\rangle$ appearing in
Eq.(\ref{eq:gamma_s}) for $\gamma_s$, the wave functions are needed:
the Schr\"odinger equation is solved with the Mapped Fourier Grid
Hamiltonian (MFGH) method \cite{kokoouline1999,pellegrini2008}. All
calculations are performed at a collision energy of 500~$\mu$K: No
large variations of the results are expected for slightly varying
collision energies, so the experimental collision energy is very
well approximated by this value. The components of the initial
radial wave function are drawn in figure~\ref{fig:wf_mf=5} in the
atomic (left column) and molecular (right column) basis. It is
striking to see in both columns that, in contrast with the generally
accepted picture of such a coupling case, the components exhibit
clear irregular features resulting from the hyperfine coupling even
at short internuclear distances, {\it i.e.} well inside the region
where the atomic states decouple into triplet and singlet states. By
looking at the molecular basis decomposition (right column), it is
clear that the singlet component (upper panel) is so strongly
coupled to the triplet component (\textit{e.g.} third panel from the
top) that they contaminate each other with their amplitude
variation. In particular, the amplitude of the singlet component is
strongly enhanced in the range of the triplet potential, {\it i.e.}
from $R=7\,a.u.$ towards large distances, compared to the uncoupled
singlet wave function (see figure \ref{fig:wf_mf=5-B}). This
``echo'' of the triplet component appearing on the singlet
projection of the wave function indicates that bound levels of the
closed channels are energetically close to the open entrance
channel. Only this situation explains the strong local perturbation
of the scattering wave function. Such a hyperfine-mediated coupling
between a bound level and a zero-energy continuum state is generally
referred to as a Feshbach resonance.

%%%%%%%%%%%%%%%%%%%%%%%%%%%%%%%%%%%%%%%%%%%%%%%%%%%%%%%%%%%%%%%%%%%%%%%%%%%%%%
\begin{figure}[thb]
\begin{center}
\includegraphics[width=0.6\textwidth,clip]{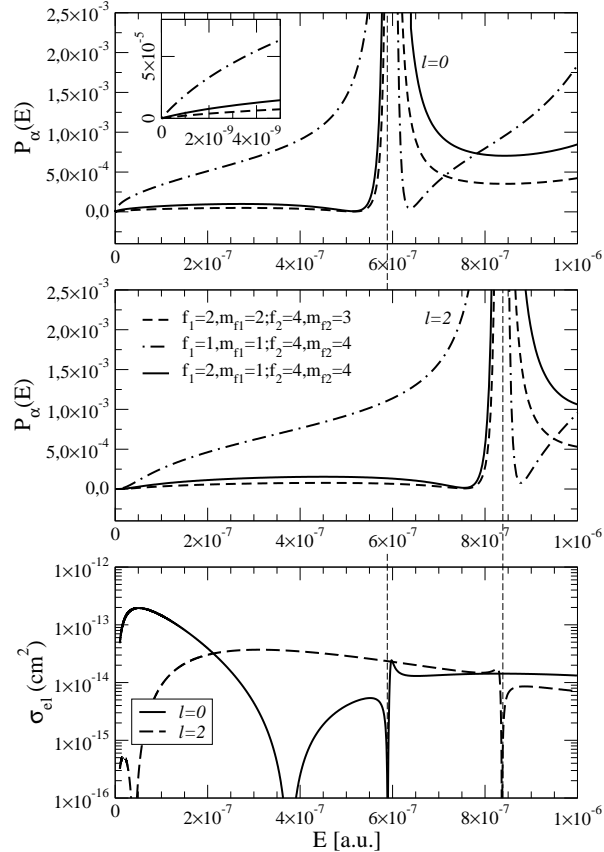}
\end{center}
\caption{\label{fig:projection} Amplitude of the $m_F$=5 initial
wave function on the three closed channels as a function of
collisional energy, for $\ell$=0 (upper panel) and $\ell$=2 (middle
panel). The inset in the upper graph shows an enlargement of the
energy range for $\ell$=0 including the collision energy of
500~$\mu$K (1.6$\times$10$^{-9}$\,a.u.), used in all other
calculations. The amplitude is given as P$_\alpha$(E), the weight of
the component $\alpha$ associated to one hyperfine channel in the
total wave function. Details are given in the text. Lower panel:
elastic cross section for the same cases $m_F$=5, $\ell$=0,2. The
positions of Feshbach resonances are indicated by vertical dashed
lines. }
\end{figure}
%%%%%%%%%%%%%%%%%%%%%%%%%%%%%%%%%%%%%%%%%%%%%%%%%%%%%%%%%%%%%%%%%%%%%%%%%%%%%%

In order to analyse how close the modeled situation is to the
Feshbach resonance, we look at the squared amplitude of the
collisional wave function on the closed channels as a function of
the scattering energy. With the total state
$\left|\Psi_{\epsilon,\ell}\right\rangle$ of the coupled system at
energy $\epsilon$ being expressed as
$\left|\Psi_{\epsilon,\ell}\right\rangle\,=\,\sum_{\alpha=1}^{4}
\psi_{\epsilon,\alpha}(R)\left|\alpha\right>$, where $\alpha$ labels
the atomic hyperfine states, we define the amplitude $P_{\alpha}(E)$
of a given hyperfine state at energy $\epsilon$ as $P_{\alpha}(E)=
\int_0^{\infty}|\psi_{\epsilon,\alpha}(R)|^2dR$. This definition is
only meaningful for closed channels, where $P_{\alpha}$=0
corresponds to an unpopulated channel and $P_{\alpha}$=1 indicates,
that $\left|\Psi_{\epsilon,\ell}\right\rangle$ is a pure bound state
in channel $\alpha$. We plot the results for the three closed
channels in figure~\ref{fig:projection}. We see that indeed a
resonance is present in the upper panel around $6\times
10^{-7}$\,a.u. (for $\ell=0$) and around $8\times 10^{-7}$\,a.u.
(for $\ell=2$), as maxima in the amplitude of the closed-channel
wave functions. Traces of these resonances are still present close
to zero collision energy, where the experiment takes place.
Therefore, the enhancement of the amplitude of the total scattering
wave function is induced by the increased mixing arising from the
presence of a broad Feshbach resonance at higher energy. In
particular the component of the $f_1=1,m_{f_1}=1,f_2=4,m_{f_2}=4$
state dominates at low energies. Note that the exact position of the
resonance might not be too reliable due to uncertainties especially
in the used triplet ground state potential, as discussed by the
authors of reference~\cite{staanum2007}. However as we will show
later, even the very weak mixing introduced by the Feshbach
resonance in this calculation leads to a convincing reproduction of
the experimentally observed rates, so the precise position of the
resonance seems of minor relevance.

We also calculate elastic cross sections for all $m_F$ subsystems
using the log-derivative algorithm of Johnson \cite{johnson1973}.
This is another way to locate and identify possible resonances
either due to the shape of potentials (shape resonances), or due to
the coupling with close-by bound states (Feshbach resonances). In
figure~\ref{fig:projection} (lowest panel) we show the
energy-dependent elastic cross section for $m_F$=5. Also here,
superimposed on the regular nodal structure of the elastic cross
section as one would expect it for an uncoupled case, the trace of
the above mentioned resonances is visible. It is important to note
that this Feshbach resonance is actually present at about the same
energy in the elastic cross-sections for all values of $m_F$. This
suggests that the singlet component of the wave function will be
perturbed for all $m_F$ in a similar way as in the presented $m_F=5$
case. Therefore the enhancement of the amplitude in the singlet
channel should be observable for all values of $m_F$ in a similar
way and we expect that the PA rate averaged over all experimentally
possible values for $m_F$ will not strongly deviate from the value
derived here for $m_F$=5.

\subsection{Comparison with experimental photoassociation rate constants}

\begin{figure}[bht]
\begin{center}
\includegraphics[width=0.7\textwidth,clip]{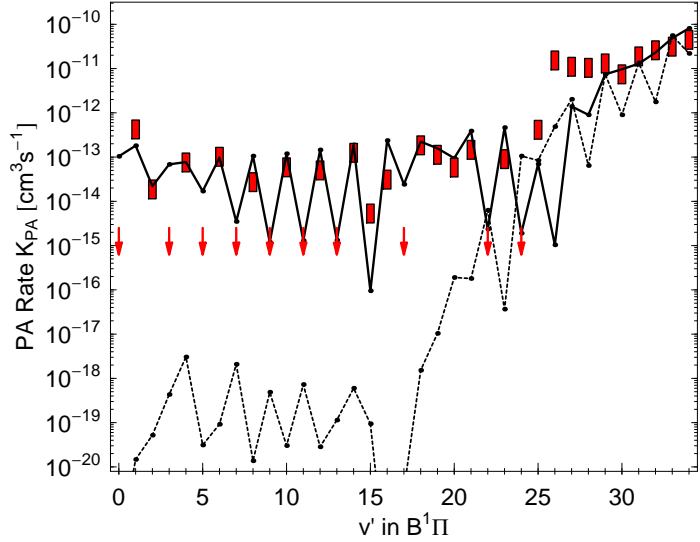}
\end{center}
\caption{\label{fig:parate_norm_log} Experimental photoassociation
rate constants (same data as figure~\ref{fig:pastrength}) for the
$J'$=1 lines (filled squares), compared to the predicted values
obtained for $m_F=5$, in the coupled-channel picture (solid line)
and in the uncoupled case (dashed line). Vibrational levels, which
could not be detected experimentally, are marked by arrows
indicating the detection threshold of the experiment. In the model
calculation, a PA laser intensity of 30\,W/cm$^2$ and a collision
energy of 500\,$\mu$K was used, corresponding to the experimental
parameters. Note that the systematic error on the experimental
values is about one order of magnitude, as described in
Sect.~\ref{sec:exp_form_rate}.}
\end{figure}

The PA rate is proportional to the transition dipole moment matrix
element between the singlet component of the initial continuum state
with the vibrational levels of the $B$ state. The calculated rates
assuming identical contributions from all $m_F$ subsystems is
represented in figure~\ref{fig:parate_norm_log} for both coupled and
uncoupled initial wave functions. As expected, both cases yield
similar rates for high vibrational states, induced by a similar
long-range part of the wave function. For low levels of the $B$
state, however, the PA rate is strongly enhanced in the
coupled-channel picture due to the ``echo'' of the triplet-like wave
function on the singlet component of the wave function. It is
evident that the oscillatory pattern of the experimental PA rate is
well reproduced by the coupled-channel model. This oscillatory
pattern arises from the variation of the overlap between the initial
scattering wave function and the wave function of the excited level
as the outer lobe of the latter ``scans'' with changing $v'$ over
the nodal structure of the former. Fair agreement between the
absolute rates from experiment and theory is also found. Even though
the measured values suffer from a systematic uncertainty of roughly
one order of magnitude, it is obvious that the experimental
observations are not compatible with the conventional picture of
uncoupled singlet and triplet states at short internuclear
distances.

%%%%%%%%%%%%%%%%%%%%%%%%%%%%%%%%%%%%%%%%%%%%%%%%%%%%%%%%%%%%%%%%%%%%%%%%%%%%%%
%%%%%%%%%%%%%%%%%%%%%%%%%%%%%%%%%%%%%%%%%%%%%%%%%%%%%%%%%%%%%%%%%%%%%%%%%%%%%%

\section{Conclusion}
\label{sec:conclusion}

By analyzing photoassociation rates of LiCs molecules in the
$B^1\Pi$ state we find that the formation rate of ground state
molecules is governed by a perturbation of the ground state
scattering wave function. The perturbation is caused by the
existence of a Feshbach resonance at low energies, which enhances
the PA rate drastically. Specifically it was shown that the singlet
component of the scattering wave function, which is responsible for
the efficient transfer into the $X^1\Sigma^+$ ground state state
through photoassociation and subsequent spontaneous emission,
contains a strong ``echo'' of the triplet wave function, which
allows for an efficient transfer of population from free pairs into
bound molecules. As the photoassociation rate into the strongly
bound levels of the excited $B^1\Pi$ state is the rate limiting
process for the formation of deeply bound ground state molecules,
the perturbation leads to an increase of the formation rate by
several orders of magnitude when compared to the uncoupled case.

To which extent our observations are specific to the LiCs case
remains to be investigated in future experiments, but we speculate
that a similar enhancement can be found in all other alkali systems.
This is supported by the fact that the observed strong enhancement
of the molecular formation rate is induced by the presence of a
Feshbach resonance roughly 200mK (or 4\,GHz) above the experimental
collision energy, a situation which is not uncommon in alkali
mixtures. We note that this kind of enhancement, or R-transfer,
through coupling in the scattering wave function is complementary to
other proposed and realized schemes for the transfer of weakly bound
triplet molecules into ground state singlet
molecules~\cite{stwalley2004}. In these schemes, the ``transfer''
from triplet to singlet character occurs via spin-orbit mixing in
the electronically excited state. In RbCs and KRb, levels in the
spin-orbit coupled states c$^3\Sigma^+$ and B$^1\Pi$ were used to
transfer weakly bound triplet molecules into absolute ground state
molecules with pulsed laser fields~\cite{sage2005,ni2008}, while in
Cs$_2$ the coupled state 0$^+_u$(b$^3\Pi_u$-A$^1\Sigma^+_u$) was
successfully employed in a STIRAP step towards deeply bound
molecules~\cite{danzl2008}. We suggest that also the wave function
of most so called Feshbach molecules might exhibit an enhanced
amplitude in the singlet channel at the inner turning point of the
triplet potential. This would facilitate the transfer of these
molecules into absolute ground state molecules via suitable excited
singlet states.

The observed strong enhancement of the PA rates into the rotational
component $J'$=2 as compared to the component $J'$=1 for low $v$'s
is still an open question which requires further investigations. A
possible explanation would be an enhancement due to a $d$-wave shape
resonance~\cite{deiglmayr2008b}, while no indication for such a
resonance was found in the model presented here. Another reason for
the observed ratio between rotational states could be rotational
coupling with higher partial waves, which was neglected in our
model. In fact, the strongly differing widths for $J'$=1 and $J'$=2
lines could indeed be an indication of the necessity to take into
account such rotational coupling at short distances.

%Acknowledgements
\ack
The experimental work is supported by the DFG under WE2661/6-1
in the framework of the Collaborative Research Project QuDipMol
within the ESF EUROCORES EuroQUAM program. JD acknowledges partial
support of the French-German University. AG is a postdoctoral fellow
of the Alexander von Humboldt-Foundation. PP and RC acknowledge
partial support from the U.S. Department of Energy, Office of Basic
Energy Sciences.

\vspace{1cm}
\section*{References}
\bibliography{mixtures}

\begin{thebibliography}{1}

\bibitem{deiglmayr2009}
J.~Deiglmayr, P.~Pellegrini, A.~Grochola, M.~Repp, R.~C\^ot\'e, O.~Dulieu, R.~Wester, and M.~Weidem\"uller.
\newblock Influence of a feshbach resonance on the photoassociation of lics.
\newblock {\em New J. Phys.}, 11:055034 (2009).

\end{thebibliography}


\begin{thebibliography}{10}

\bibitem{doyle2004}
J.~Doyle, B.~Friedrich, R.~V. Krems, and F.~Masnou-Seeuws.
\newblock {\em Eur. Phys. J. D}, 31:149, 2004.

\bibitem{dulieu2006}
O.~Dulieu, M.~Raoult, and E.~Tiemann.
\newblock {\em J. Phys. B}, 39(19), 2006.
\newblock Introductory review for the special issue on {Cold Molecules}.

\bibitem{zelevinsky2008}
T.~Zelevinsky, S.~Kotochigova, and J.~Ye.
\newblock {\em Phys. Rev. Lett.}, 100:043201, 2008.

\bibitem{tscherbul2006}
T.~V. Tscherbul and R.~V. Krems.
\newblock {\em Phys. Rev. Lett.}, 97:083201, 2006.

\bibitem{micheli2006}
A.~Micheli, G.~K. Brennen, and P.~Zoller.
\newblock {\em Nature Physics}, 2:341, 2006.

\bibitem{pupillo2008}
G.~Pupillo, A.~Griessner, A.~Micheli, M.~Ortner, D.-W. Wang, and P.~Zoller.
\newblock {\em Phys. Rev. Lett.}, 100:050402, 2008.

\bibitem{DeMille2002}
D.~DeMille.
\newblock {\em Phys. Rev. Lett.}, 88:067901, 2002.

\bibitem{yelin2006}
S.~F. Yelin, K.~Kirby, and R.~{C\^ot\'e}.
\newblock {\em Phys. Rev. A}, 74:050301, 2006.

\bibitem{rabl2006}
P.~Rabl, D.~DeMille, J.~M. Doyle, M.~D. Lukin, R.~J. Schoellkopf, and
  P.~Zoller.
\newblock {\em Phys. Rev. Lett.}, 97:033003, 2006.

\bibitem{charron2007}
E.~Charron, P.~Milman, A.~Keller, and O.~Atabek.
\newblock {\em Phys. Rev. A}, 75:033414, 2007.

\bibitem{lang2008}
F.~Lang, K.~Winkler, C.~Strauss, R.~Grimm, and J.~Hecker Denschlag.
\newblock {\em Phys. Rev. Lett.}, 101:133005, 2008.

\bibitem{danzl2008}
J.~G. Danzl, E.~Haller, M.~Gustavsson, M.~J. Mark, R.~Hart, N.~Bouloufa,
  O.~Dulieu, H.~Ritsch, and H.-C. N{\"a}gerl.
\newblock {\em Science}, 321:1062, 2008.
\newblock H.-C. N{\"a}gerl \textit{private communication}.

\bibitem{danzl2008b}
J.~G. Danzl, M.~J. Mark, E.~Haller, M.~Gustavsson, N.~Bouloufa, O.~Dulieu,
  H.~Ritsch, R.~Hart, and H.-C. Naegerl.
\newblock {\em arXiv:0811.2374v1 [cond-mat.other]}, 2008.

\bibitem{ni2008}
K.-K. Ni, S.~Ospelkaus, M.~H.~G. de~Miranda, A.~Pe'er, B.~Neyenhuis, J.~J.
  Zirbel, S.~Kotochigova, P.~S. Julienne, D.~S. Jin, and J.~Ye.
\newblock {\em Science}, 322:231, 2008.

\bibitem{nikolov2000}
A.~N. Nikolov, J.~R. Ensher, E.~E. Eyler, H.~Wang, W.~C. Stwalley, and P.~L.
  Gould.
\newblock {\em Phys. Rev. Lett.}, 84:246, 2000.

\bibitem{sage2005}
J.~M. Sage, S.~Sainis, T.~Bergeman, and D.~{DeMille}.
\newblock {\em Phys. Rev. Lett.}, 94:203001, 2005.

\bibitem{viteau2008}
M.~Viteau, A.~Chotia, M.~Allegrini, N.~Bouloufa, O.~Dulieu, D.~Comparat, and
  P.~Pillet.
\newblock {\em Science}, 321:232, 2008.

\bibitem{deiglmayr2008b}
J.~Deiglmayr, A.~Grochola, M.~Repp, K.~M{\"o}rtlbauer, C.~Gl{\"u}ck, J.~Lange,
  O.~Dulieu, R.~Wester, and M.~Weidem{\"u}ller.
\newblock {\em Phys. Rev. Lett.}, 101:133004, 2008.

\bibitem{aymar2005}
M.~Aymar and O.~Dulieu.
\newblock {\em J. Chem. Phys.}, 122:204302, 2005.

\bibitem{deiglmayr2009a}
J.~Deiglmayr, M.~Repp, A.~Grochola, K.~M{\"o}rtlbauer, C.~Gl{\"u}ck, J.~Lange,
  O.~Dulieu, R.~Wester, and M.~Weidem{\"u}ller.
\newblock {\em Accepted for publication in {Faraday Discussions} \textbf{142}},
  2009.
\newblock Also at arXiv:0812.1002v1.

\bibitem{pellegrini2008}
P.~Pellegrini, M.~Gacesa, and R.~{C\^ot\'e}.
\newblock {\em Phys. Rev. Lett.}, 101:053201, 2008.

\bibitem{kraft2007}
S.~D. Kraft, J.~Mikosch, P.~Staanum, J.~Deiglmayr, J.~Lange, A.~Fioretti,
  R.~Wester, and M.~Weidem\"uller.
\newblock {\em Appl. Phys. B}, 89:453, 2007.

\bibitem{ketterle1993}
W.~Ketterle, K.~B. Davis, M.~A. Joffe, A.~Martin, and D.~E. Pritchard.
\newblock {\em Phys. Rev. Lett.}, 70:2253, 1993.

\bibitem{jones1999}
K.~M. Jones, P.~D. Lett, E.~Tiesinga, and P.~S. Julienne.
\newblock {\em Phys. Rev. A}, 61:012501, 1999.

\bibitem{thorsheim1987}
H.~R. Thorsheim, J.~Weiner, and P.S. Julienne.
\newblock {\em Phys. Rev. Lett.}, 58:2420, 1987.

\bibitem{jones2006}
K.~M. Jones, E.~Tiesinga, P.~D. Lett, and P.~S. Julienne.
\newblock {\em Rev. Mod. Phys.}, 78:483, 2006.

\bibitem{cote1997}
R.~{C\^ot\'e} and A.~Dalgarno.
\newblock {\em Chem. Phys. Lett.}, 279:50, 1997.

\bibitem{stein2008}
A.~Stein, A.~Pashov, P.F. Staanum, H.~Kn\"ockel, and E.~Tiemann.
\newblock {\em Eur. Phys. J. D}, 48:177, 2008.

\bibitem{verner1996}
D.~A. Verner, G.~J. Ferland, K.~T. Korista, and D.~G. Yakovlev.
\newblock {\em Astrophys. J.}, 465:487, 1996.

\bibitem{fraser2002}
G.~W. Fraser.
\newblock {\em Int. J. Mass Spect.}, 215:13, 2002.

\bibitem{drag2000}
C.~Drag, B.~Laburthe Tolra, O.~Dulieu, D.~Comparat, M.~Vatasescu, S.~Boussen,
  S.~Guibal, A.~Crubellier, and P.~Pillet.
\newblock {\em IEEE J. Quantum Electron.}, 36:1378, 2000.

\bibitem{wester2004}
R.~Wester, S.D. Kraft, M.~Mudrich, M.U. Staudt, J.~Lange, N.~Vanhaecke,
  O.~Dulieu, and M.~Weidem{\"u}ller.
\newblock {\em Appl. Phys. B}, 79:993, 2004.

\bibitem{mancini2004a}
M.~W. Mancini, G.~D. Telles, A.~R.~L. Caires, V.~S. Bagnato, and L.~G.
  Marcassa.
\newblock {\em Phys. Rev. Lett.}, 92:133203, 2004.

\bibitem{Azizi2004}
S.~Azizi, M.~Aymar, and O.~Dulieu.
\newblock {\em Eur. J. Phys. D}, 31:195, 2004.

\bibitem{napolitano1994}
R.~Napolitano, J.~Weiner, C.~J. Williams, and P.~S. Julienne.
\newblock {\em Phys. Rev. Lett.}, 73:1352, 1994.

\bibitem{inprepA}
A.~Grochola, A.~Pashov, J.~Deiglmayr, M.~Repp, E.~Tiemann, R.~Wester, and
  M.~Weidem\"uller.
\newblock \textit{In preparation}.

\bibitem{bergeman2002}
T.~Bergeman, P.~S. Julienne, C.~J. Williams, E.~Tiesinga, M.~R. Manaa, H.~Wang,
  P.~L. Gould, and W.~C. Stwalley.
\newblock {\em J. Chem. Phys.}, 117:7491, 2002.

\bibitem{udem1999}
Th. Udem, J.~Reichert, R.~Holzwarth, and T.~W. H{\"a}nsch.
\newblock {\em Phys. Rev. Lett.}, 82:3568, 1999.

\bibitem{udem2000}
Th. Udem, J.~Reichert, T.~W. H{\"a}nsch, and M.~Kourogi.
\newblock {\em Phys. Rev. A}, 62:031801, 2000.

\bibitem{kokoouline1999}
V.~Kokoouline, O.~Dulieu, R.~Kosloff, and F.~Masnou-Seeuws.
\newblock {\em J. Chem. Phys.}, 110:9865, 1999.

\bibitem{staanum2007}
P.~Staanum, A.~Pashov, H.~Kn{\"o}ckel, and E.~Tiemann.
\newblock {\em Phys. Rev. A}, 75:042513, 2007.

\bibitem{johnson1973}
B.R. Johnson.
\newblock {\em J. Comp. Phys.}, 13:445, 1973.

\bibitem{stwalley2004}
W.C. Stwalley.
\newblock {\em Eur. Phys. J. D}, 31:221, 2004.

\end{thebibliography}
\end{document}

\title{Erratum on the paper: "Influence of a Feshbach resonance on the photoassociation of LiCs", New Journal of Physics, vol. 11, p.055034 (2009).}

\author{J Deiglmayr, A Grochola, M Repp, R Wester, M Weidem\"uller\footnote{\ead{weidemueller@physi.uni-heidelberg.de}},}
\address{Physikalisches Institut, Albert-Ludwigs-Universit\"at Freiburg, Germany}
\author{O Dulieu,}
\address{Laboratoire Aim\'e Cotton, CNRS, University Paris-Sud XI, Orsay, France}
\author{P Pellegrini and R C\^ot\'e}
\address{Department of Physics, University of Connecticut, Storrs, USA}

\begin{abstract}
Erratum on the paper: "Influence of a Feshbach resonance on the photoassociation of LiCs", New Journal of Physics, vol. 11, p.055034 (2009).
\end{abstract}

\pacs{ 37.10.Mn, 33.20.-t, 33.80.Rv}
\maketitle

In the paper by Deiglmayr {\it et al} \cite{deiglmayr2009}, we found a sign error in the transformation of the scattering wave function between the atomic basis (eq.~6) and the molecular basis (eq.~7) that we used for the case $m_F=5$ for the following atomic state:

\begin{eqnarray*}
&|f_1=2,m_{f_1}  = 2,f_2=3,m_{f_2}=3 \rangle =  \\
& + \sqrt{\frac{7}{16}}|S=0, m_S=0, I=5, m_I=5 \rangle  \\
& + \sqrt{\frac{7}{16}}|S=1, m_S=0, I=5, m_I=5 \rangle \\
& - \sqrt{\frac{3}{80}}|S=1, m_S=1, I=4, m_I=4 \rangle \\
& - \sqrt{\frac{7}{80}} |S=1, m_S=1, I=5, m_I=4 \rangle
\end{eqnarray*}

In consequence, the results of the left column of Fig.~5 of the quoted paper are correct, while those of the right column are not. The corrected figure is now displayed in Figure \ref{fig:newfig5} below. In particular there is not anymore influence of the triplet part of the wave function on the singlet one (see upper panel of the right column). This affects also Figures 6 and 7 of reference~\cite{deiglmayr2009}. The black trace of the upper part of Figure 6 is wrong, as well as the results of Figure 7.

The main consequence concerns the interpretation of the experimental results reported in Figure 8 of Deiglmayr {\it et al} \cite{deiglmayr2009}. The magnitude and modulation of the photoassociation (PA) rate observed in the experiment can be interpreted only if the triplet component of the scattering wave function is involved, as it was discussed in the paper. However, the model employed in reference~\cite{deiglmayr2009} cannot reproduce these results.

The role of the triplet component of the scattering wave function can be invoked in the PA towards the $B^1\Pi$ levels by considering the small spin-orbit coupling of the $B^1\Pi$ state with the neighboring triplet states (see Figure 4 of Deiglmayr {\it et al} \cite{deiglmayr2009}), while the modulations of the PA rate may be influenced by a slight variation of the scattering length of the lowest $a^3\Sigma^+$ state. These issues will be considered in a future paper.

We would like to thank Prof. E. Tiemann for pointing out to us this error sign.

%%%%%%%%%%%%%%%%%%%%%%%%%%%%%%%%%%%%%%%%%%%%%%%%%%%%%%%%%%%%%%%%%%%%%%%%%%%%%%
\begin{figure}
\begin{center}
\includegraphics[width=0.6\textwidth,clip]{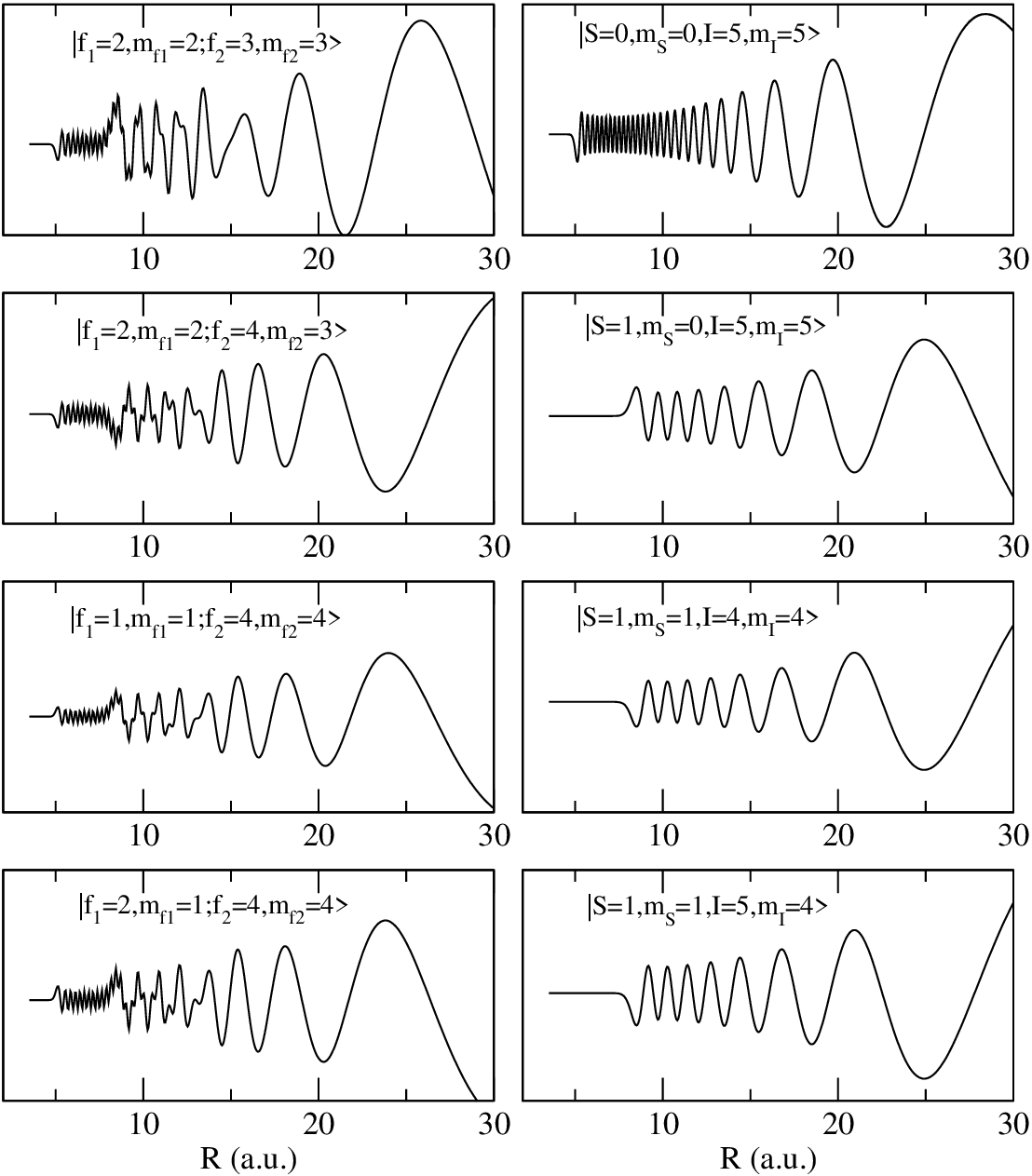}
\end{center}
\caption{\label{fig:newfig5} Components of the energy-normalized radial wave function for a Li–Cs pair colliding at 500~$\mu$K, in the atomic basis (left column) and in the molecular basis (right column). The displayed case corresponds to $m_{f_1}$+$m_{f_2}$=$5$. Under the given experimental conditions, the upper-most atomic state $f_1 = 2$, $m_{f_1} = 2$, $f_2 = 3$, $m_{f_2} = 3$ is the only open entrance channel, the three other shown states are closed channels. Note that there is no line-to-line relation between the states listed in each column.}
\end{figure}
%%%%%%%%%%%%%%%%%%%%%%%%%%%%%%%%%%%%%%%%%%%%%%%%%%%%%%%%%%%%%%%%%%%%%%%%%%%%%%

%\bibliography{mixtures}